\documentclass[singlespacing]{elsart}

\usepackage{graphicx}

\usepackage{amssymb}
\journal{}
\begin{document}

\begin{frontmatter}

\title{Exact solution of Schr\"odinger equation with symmetric double-well potential versus WKB: accuracy for ground state splitting}

\author{ A.E. Sitnitsky},
\ead{sitnitsky@mail.kibb.knc.ru}

\address{Kazan Institute of Biochemistry and Biophysics, P.O.B. 30, Kazan
420111, Russia. Tel. 8-843-231-90-37. e-mail: sitnitsky@kibb.knc.ru }

\begin{abstract}
The one-dimensional Schr\"odinger equation with symmetric trigonometric double-well potential (DWP) is exactly solved via angular oblate spheroidal function. The results of stringent analytic calculation for the ground state splitting of hydrogen bond in malonaldehyde are compared with several variants of approximate semiclassical (WKB) ones. This enables us to compare the accuracy of various WKB formulas suggested in the literature: 1. ordinary WKB, i.e., the formula from the Landau and Lifshitz textbook; 2. Garg's formula; 3. instanton approach. The results obtained provide a new theoretical tool for the precise quantitative description of experimental data on IR spectroscopy of malonaldehyde.
\end{abstract}

\end{frontmatter}

\section{Introduction}
Schr\"odinger equation (SE) with a double well potential (DWP) is widely applied in physics and chemistry. The most well-known example is
the inversion of ammonia molecule $NH_3$ \cite{Ros32}, \cite{Man35} that played a cornerstone role in the development of radiospectroscopy and quantum electronics as the basis for the first maser \cite{Tow55}. More recent applications include heterostructures, Bose-Einstein condensates and superconducting circuits (see \cite{Xie12}, \cite{Dow13}, \cite{Che13}, \cite{Har14}, \cite{Dow16}, \cite{Dow17}, \cite{Tur10}, \cite{Tur16} and refs. therein). We consider here only smooth DWP and pass over numerous models with rectangular wells or sewing together two single-well potentials (harmonic, Morse, etc.). Although the latter are very useful in revealing the underlying physics in many model systems pertaining in particular to semiconducting devices or optronics their accuracy is always under question. In contrast smooth potentials provide mathematically rigorous treatment of the problem under consideration and precise description of relevant experimental data. Their drawback is that one has to deal with rather complex special functions of mathematical physics. The above mentioned single-well potentials have exhausted the ability of habitual hypergeometric function to represent the solution of SE. For DWP we inevitably have to resort to more complicated and less familiar constructions such as confluent Heun's function (CHF), spheroidal function (SF) or Coulomb (generalized) SF.
Fortunately considerable progress in their implementations in modern mathematical software packages such as {\sl {Maple}} or {\sl {Mathematica}} makes their usage to be a routine procedure.

One can witness noticeable progress in obtaining quasi-exact (i.e., exact for some particular choice of potential parameters) \cite{Xie12}, \cite{Dow13}, \cite{Che13} and exact (those for an arbitrary set of potential parameters) \cite{Har14}, \cite{Sit17}, \cite{Sit171} solutions for SE with DWP by their reducing to the confluent Heun's equation (CHE). A plenty of potentials for SE are shown to be exactly solvable via CHF \cite{Ish16}. The latter is a well studied special function tabulated in {\sl {Maple}} \cite{Fiz12}, \cite{Fiz10}, \cite{Sha12}. As a result the obtained solution of SE very convenient for usage. In recent years the exact solution of the Smoluchowski equation for reorientational motion in Maier-Saupe DWP was obtained via CHF \cite{Sit15}, \cite{Sit16}. The approach yields the probability distribution function in the form convenient for application to nuclear spin-lattice relaxation \cite{Sit11}. In the present article we apply similar technique to SE with trigonometric DWP.

On the other hand one of the main tools for investigating SE is the famous semiclassical approximation (WKB method) in both the ordinary variant \cite{Lan74}, \cite{Gar00}, \cite{Par97}, \cite{Par98}, \cite{Son08}, \cite{Ras12}, \cite{Son15} and the instanton approach \cite{Col85}, \cite{Kle95}, \cite{Tur10}, \cite{Tur16}, \cite{Gil77}, \cite{Neu78}. By present it has attained impressive maturity at the same time remaining an active field of researches. Thus it seems interesting to compare the results of various WKB formulas with those of exact solution for some model potential. For this purpose we choose the trigonometric DWP \cite{Sit17}, \cite{Sit171}. The latter is a particular case of some general potential from \cite{Ish16} (N2 with $m_{1,2}=\left(1/2,1/2\right)$ from Table.1). For this DWP the wave functions can be expressed via confluent Heun's function (CHF) or equivalently via Coulomb (generalized) SF. For the case of symmetric DWP the latter is reduced to angular oblate SF that is realized in {\sl {Mathematica}} and as a result is very convenient for application. Also its spectrum of eigenvalues is realized in {\sl {Mathematica}} that makes the calculation of energy levels for trigonometric DWP to be an instant (at a click) procedure. Thus the obtained exact solution of SE with this potential provides considerable facilities and can be a reliable referee point for comparison of the accuracy of various WKB formulas \cite{Sit171}. For the case of asymmetric DWP the variant with CHF is more convenient at present. The reason for this is the fact that Coulomb (generalized) SF up to now has not been implemented as the standard package into Mathematica although such package was developed as early as in 2001 by Falloon \cite{Fal01}. So long as we are concerned with the structure of energy levels (in particular with the ground state splitting) CHF realization in {\sl {Maple}} works well. Only when we deal with calculating the integrals with the wave function there arises a problem with this realization \cite{Sit17}, \cite{Sit171} and one has to circumvent it somehow.

The aim of the article is to solve exactly SE with the trigonometric DWP, to find energy levels and to provide analytic representation of the wave functions for them. We show that the latter are expressed via SF. We exemplify our general results by detailed treating the hydrogen bond in malonaldehyde (experimental data are taken from \cite{Fil05}) that is a prototype example of symmetric DWP. For this object detailed data on IR spectroscopy are available along with those of large scattering cross-section of protons (see \cite{Fil05} and refs. therein). Precise experimental data for tunneling phenomena in malonaldehyde make it to be ideal test object for verifying the accuracy of various theoretical methods. The main aim of the article is to compare the obtained exact calculation for the ground state splitting in malonaldehyde with those of several WKB formulas. We provide detailed comparison of our stringent result with those of ordinary WKB approach (Landau and Lifshitz textbook formula for the ground state splitting in a symmetric potential \cite{Lan74}), Garg's formula \cite{Gar00} and the instanton approach \cite{Col85}, \cite{Kle95}.

In closing this Sec. it seems expedient to add a comment on the chosen object for our study. IR spectroscopy of malonaldehyde (see \cite{Fil05}, \cite{Yan10}, \cite{Wu16}, \cite{Wu161}, \cite{Wan08}, \cite{Fir91}, \cite{Bab99}, \cite{Car84}, \cite{Vie07} and refs. therein) is a subject of nondecreasing intensive activity. In \cite{Car84} the authors concluded that if the reaction path in a polyatomic molecule is sharply curved then it is preferentially to avoid the one-dimensional potential energy surface in favor of two coordinates. Further  they applied their arguments to malonaldehyde \cite{Car86} and since then the point of view became widely accepted that "the proton tunneling in malonaldehyde cannot be reduced to a one-dimensional problem, in other words, the large amplitude motion is not restricted to a single reaction coordinate" \cite{Vie07}. Up to now it was widely believed that the precise details of malonaldehyde IR spectrum (ground state splitting in particular) can not be correctly reproduced in a one-dimensional model. In spite of this attitude one-dimensional models remain to be a useful tool in investigating malonaldehyde \cite{Fil05}. In \cite{Fil05} numerical solution of SE was carried out for a model DWP. In the present paper we obtain analytic solution of the problem for the trigonometric DWP. The results obtained lead to equally good agreement with experimental data as those of numerical calculations \cite{Fil05} and besides provide additional facilities in treating the problem. The numerical \cite{Fil05} and analytic (present article) approaches testify that one-dimensional models are able to provide excellent accuracy in describing the energy levels structure of malonaldehyde and precise details of its IR spectrum.

The paper is organized as follows.  In Sec. 2 the problem under study is formulated.  In Sec. 3 the solution of SE via SF is presented. The next sections are devoted to comparing the exact result with those of different WKB approaches to the ground state splitting. In Sec. 4 the formula from Landau and Lifshitz textbook is analyzed. In Sec.5 the Garg's formula is investigated. In Sec.6 the instanton approach is studied. In Sec.7 the results are discussed and the conclusions are summarized.

\section{Schr\"odinger equation and the potential}
We consider the one-dimensional SE
\begin{equation}
\label{eq1} \frac{d^2 \psi (x)}{dx^2}+\frac{2M}{\hbar^2}\left[E-V(x)\right]\psi (x)=0
\end{equation}
where $V(x)$ is a DWP that is infinite at the boundaries of the finite interval $x=\pm L$.
We introduce the dimensionless energy $\epsilon$, the dimensionless distance $y$ and the dimensionless potential $U(y)$
\begin{equation}
\label{eq2} \epsilon=\frac{8ML^2E}{\hbar^2 \pi^2}\ \ \ \ \ \ \ \ \ \ \ \ \ y=\frac{\pi x}{2L}\ \ \ \ \ \ \ \ \ \ \ \ \ \ \ \ \ U(y)=\frac{8ML^2}{\hbar^2 \pi^2}V(x)
\end{equation}
so that $-\pi/2\leq y \leq \pi/2$. The We choose trigonometric DWP
\begin{equation}
\label{eq3} U(y)=h\ \tan^2 y-b\sin^2 y+a\sin y
\end{equation}
For the symmetric case ($a=0$) of the trigonometric DWP the barrier height $B=-U\left(y_{min}\right)$ and barrier width $\Delta=y_{min}^{(1)}-y_{min}^{(2)}$ are related with the parameters $h$ and $b$ as follows
\[
\Delta=2\arccos\left(\frac{h}{b}\right)^{1/4}\ \ \ \ \ \ \ \ \ \ \ \ \ \ \ \ \ \ \ \ \ \ \ \ B=\left(\sqrt {h}-\sqrt {b}\right)^2
\]

Inversely one obtains
\[
b=\frac{B}{\left\{1-\left[\cos\left(\Delta/2\right)\right]^2\right\}^2} \ \ \ \ \ \ \ \ \ \ \ \ \ \ \ \ \ \ \ \ \ \ \ h=\frac{B\left[\cos\left(\Delta/2\right)\right]^4}{\left\{1-\left[\cos\left(\Delta/2\right)\right]^2\right\}^2}
\]
The dimensionless SE with the potential (\ref{eq3}) takes the form
\begin{equation}
\label{eq4} \psi''_{yy} (y)+\left[\epsilon-h\ \tan^2 y +b\ \sin^2 y-a\ \sin y\right]\psi (y)=0
\end{equation}

\section{Solution of the Schr\"odinger equation}
We denote
\begin{equation}
\label{eq5}h=m^2-\frac{1}{4}
\end{equation}
\begin{equation}
\label{eq6}b=p^2
\end{equation}
The example of the potential (\ref{eq3}) in these designations for a symmetric case is depicted in Fig.1.
\begin{figure}
\begin{center}
\includegraphics* [width=\textwidth] {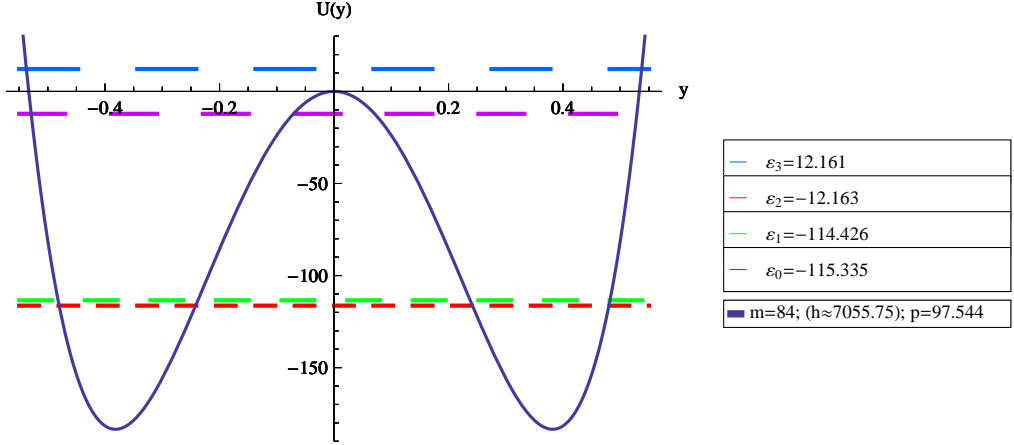}
\end{center}
\caption{The model double-well potential (\ref{eq3}) in designations (\ref{eq5}), (\ref{eq6}) at the values of the parameters $m=84$, ($h\approx 7055.75$), $p=97.544$, $a=0$. The parameters are chosen to describe the potential and the energy levels for the hydrogen bond in malonaldehyde
(experimental data are taken from \cite{Fil05}). The energy levels $\epsilon_0=-115.335$, $\epsilon_1=-114.426$, $\epsilon_2=-12.163$, $\epsilon_3=12.161$ are respectively depicted by the dashes of increasing length. The splitting of the ground state $\epsilon_1-\epsilon_0=0.909$ corresponds to $21.583\ {\rm cm^{-1}}$ in dimensional units.} \label{Fig.1}
\end{figure}
We introduce a new variable
\begin{equation}
\label{eq7} s=\sin y
\end{equation}
where $-1\leq s \leq 1$ and a new function $v(s)$ by the relationship
\begin{equation}
\label{eq8} \psi (y)=v(\sin y)\cos^{1/2} y
\end{equation}
The equation for $v(s)$ is
\begin{equation}
\label{eq9} \frac{d}{ds}\left[\left(1-s^2\right)\frac{dv(s)}{ds}\right]+\Biggl\{p^2s^2-as+
\epsilon+m^2-\frac{1}{2}-
\frac{m^2}{1-s^2}\Biggr\}v(s)=0
\end{equation}

If $m$ is integer then (\ref{eq9}) belongs to the type of Coulomb (generalized) spheroidal equations \cite{Kom76}
and its solution is
\begin{equation}
\label{eq10} v(s)=\bar\Xi_{mq}\left(p, -a;s\right)
\end{equation}
where $q=0,1,2,...$ and $\bar\Xi_{mq}\left(p, -a;s\right)$ is CSF. The energy levels are determined from the relationship
\begin{equation}
\label{eq11} \epsilon_q=\lambda_{mq}\left(p, -a\right)+\frac{1}{2}-p^2-m^2
\end{equation}
where $\lambda_{mq}\left(p, -a\right)$ is the spectrum of eigenvalues for $\bar\Xi_{mq}\left(p, -a;s\right)$.
As a result the wave function takes the form
\begin{equation}
\label{eq12} \psi_q (y)=\cos^{1/2} y\ \bar\Xi_{mq}\left(p, -a;\sin y\right)
\end{equation}
Unfortunately the CSF is realized neither in {\sl {Maple}} nor in the standard package of {\sl {Mathematica}} at present. The application of the solution (\ref{eq12}) requires a tedious implementation in {\sl {Mathematica}} of the corresponding package developed by Falloon \cite{Fal01}.

For the symmetric case of DWP ($a=0$) eq. (\ref{eq9}) belongs to the type of angular oblate spheroidal equation \cite{Kom76} (it can also  be considered  as the limit $ \bar \Xi_{mq}\left(p, 0;s\right)=\bar S_{m(q+m)}\left(p;s\right)$ of the above formulas)
and its solution is
\begin{equation}
\label{eq13} v(s)=\bar S_{m(q+m)}\left(p;s\right)
\end{equation}
where $q=0,1,2,...$ and $\bar S_{m(q+m)}\left(p;s\right)$ is angular oblate SF. The latter is realized in  {\sl {Mathematica}} as $\bar S_{m(q+m)}\left(p;s\right)\equiv SpheroidalPS[(q+m),m,ip,s]$. The energy levels are determined from the relationship
\begin{equation}
\label{eq14}
\epsilon_q=\lambda_{m(q+m)}\left(p\right)+\frac{1}{2}-p^2-m^2
\end{equation}
where $\lambda_{m(q+m)}\left(p\right)$ is the spectrum of eigenvalues for $\bar S_{m(q+m)}\left(p;s\right)$. It is realized in {\sl {Mathematica}} as $\lambda_{m(q+m)}\left(p\right)\equiv SpheroidalEigenvalue[(q+m),m,ip]$.
The ground state splitting is
\begin{equation}
\label{eq15} \epsilon_1-\epsilon_0=\lambda_{m(1+m)}\left(p\right)-\lambda_{mm}\left(p\right)
\end{equation}
Thus we have
\begin{equation}
\label{eq16} \psi_q (y)=\cos^{1/2} y\ \bar S_{m(q+m)}\left(p;\sin y\right)
\end{equation}

The formulas (\ref{eq16}), (\ref{eq14}) and (\ref{eq15}) provide a highly efficient and convenient tool for calculating the wave functions, the energy levels and the ground state splitting for SE with symmetric case of trigonometric DWP (\ref{eq3}) with the help of {\sl {Mathematica}}.

\section{Landau and Lifshitz textbook formula for a symmetric potential}
The most widely known expression for the ground state splitting in a symmetric DWP $V(x)$ is \cite{Lan74}
\begin{equation}
\label{eq17} E_1-E_0=\frac{\hbar\omega}{\pi} \exp\left[-\frac{\sqrt {2M}}{\hbar}\int_{-\bar c}^{\bar c}dx\ \left \vert\sqrt{V(x)-E_m}\right \vert\right]
\end{equation}
where $E_m=\left(E_1-E_0\right)/2$, $\bar c$ is the turning point corresponding to $E_m$ and $\omega$ is the classical vibration frequency for the well of $V(x)$. We cast (\ref{eq17}) into dimensionless form making use of (\ref{eq2}). For this purpose we also need the relationship of the frequency $\omega$ from the dimensional potential in the vicinity of the minimum $x_{min}$
\begin{equation}
\label{eq18} V(x)
\left| {\begin{array}{l}
  \\
{x\sim x_{min}}\\
 \end{array}}\right. \approx V\left(x_{min}\right)+\frac{M\omega^2}{2}\left(x-x_{min}\right)^2
\end{equation}
with the derivative of the dimensionless potential $U(y)$ at $y=y_{min}$. It has the form
\begin{equation}
\label{eq19} \omega=\frac{\hbar \pi^2}{4ML^2}\sqrt{\frac{1}{2}\left(\frac{d^2U(y)}{dy^2}\right)\left| {\begin{array}{l}
  \\
{y_{min}}\\
 \end{array}}\right. }
\end{equation}
We denote $c$ the dimensionless analog of $\bar c$
\begin{equation}
\label{eq20}  c=\frac{\pi \bar c}{2L}
\end{equation}
Then the dimensionless ground state splitting is
\begin{equation}
\label{eq21} \epsilon_1-\epsilon_0=\frac{2}{\pi}\left(\frac{1}{2}\left (\frac{d^2 U(y)}{dy^2}\right)\left| {\begin{array}{l}
  \\
{y_{min}}\\
 \end{array}}\right. \right)^{1/2}\exp\left(-\int_{c}^{c}dy\ \left \vert \sqrt{U(y)-\epsilon_m}\right\vert\right)
\end{equation}
For malonaldehyde $c=0.239742$ (see Fig.1). The calculation of (\ref{eq21}) yields $\epsilon_1-\epsilon_0=0.922466$ in good agreement with the exact value $\epsilon_1-\epsilon_0=0.909$.

\section{Garg's formula for a symmetric potential}
Garg \cite{Gar00} suggested an extremely useful expression (see \cite{Ras12} for a more convenient form of the Garg's formula)
\begin{equation}
\label{eq22}  E_1-E_0=\hbar\omega \left(\frac{M\omega \bar d^2}{\pi\hbar}\right)^{1/2}\exp\left(A-S_0/\hbar\right)
\end{equation}
where $\pm \bar d$ are the positions of the minima of the potential,
\begin{equation}
\label{eq23} S_0=\int_{-\bar d}^{\bar d}dx\ \sqrt{2M\left[V(x)-V\left(\bar d\right)\right]}
\end{equation}
and
\begin{equation}
\label{eq24} A=\int_{0}^{\bar d}dx\ \left[\frac{M\omega}{\sqrt{2M\left[V(x)-V\left(\bar d\right)\right]}}-\frac{1}{\bar d-x}\right]
\end{equation}
The convenience of the formula (\ref{eq22}) compared with (\ref{eq17}) is in the fact that the integration is carried out between the minima the potential rather that between the turning points corresponding to $E_m=\left(E_1-E_0\right)/2$.

We cast (\ref{eq22}) into the dimensionless form. The relationship of the frequency $\omega$ from the dimensional potential in the vicinity of the minimum $x_{min}$ is given by  (\ref{eq19}).
We denote $d$ the dimensionless analog of $\bar d$
\begin{equation}
\label{eq25} d=\frac{\pi \bar d}{2L}
\end{equation}
Then the dimensionless ground state splitting is
\[
\epsilon_1-\epsilon_0=2d\left(\frac{1}{2}\left (\frac{d^2 U(y)}{dy^2}\right)\left| {\begin{array}{l}
  \\
{y_{min}=d}\\
 \end{array}}\right. \right)^{3/4}\times
\]
\[
\exp\Biggl\{\int_{0}^{d}dy\ \Biggl[\left(\frac{1}{2}\left (\frac{d^2 U(y)}{dy^2}\right)\left| {\begin{array}{l}
  \\
{y_{min}=d}\\
 \end{array}}\right. \right)^{1/2}\times
\]
\begin{equation}
\label{eq26} \frac{1}{\sqrt{U(y)-U(d)}}-\frac{1}{d -y}\Biggr]-\int_{-d}^{d}dy\ \sqrt{U(y)-U(d)}\Biggr\}
\end{equation}
For malonaldehyde $d=0.381856$ (see Fig.1). The calculation of (\ref{eq26}) yields $\epsilon_1-\epsilon_0=0.973715$.

\section{Instanton formula for a symmetric potential}
The instanton approach \cite{Col85}, \cite{Kle95} yields for the ground state splitting the following expression
\begin{equation}
\label{eq27}  E_1-E_0=2\hbar K\left(\frac{S_0}{2\pi\hbar}\right)^{1/2}\exp\left(-S_0/\hbar\right)
\end{equation}
Here $S_0$ is given by (\ref{eq23}). The usual way to calculate the
 $K$ is as follows \cite{Col85}
\begin{equation}
\label{eq28}  K=\left[\frac{det\left(-\partial ^2_\tau+\omega^2\right)}{det'\left[-\partial ^2_\tau+M^{-1}V''\left(x_{cl}(\tau)\right)\right]}\right]^{1/2}
\end{equation}
where $det'$ indicates that the zero eigenvalue is to be omitted when computing the determinant. The instanton $x_{cl}$ obeys the classical equation of motion
\begin{equation}
\label{eq29} M\frac{d^2x_{cl}(\tau)}{d\tau^2} -V'\left(x_{cl}\right)=0
\end{equation}
which has a solution
\begin{equation}
\label{eq30} \tau=\sqrt {\frac{M}{2}}\int_{0}^{x_{cl}}\frac{dx}{\sqrt{V(x)-V\left(\bar d\right)}}
\end{equation}
with the boundary conditions $x_{cl}\left(-T/2\right)=-\bar d$ and $x_{cl}\left(T/2\right)=\bar d$.

The determinants in (\ref{eq28}) are calculated as the products of the eigenvalues of a corresponding equation. This usual way is inapplicable in our case because for our potential (\ref{eq3}) such equation can not be solved in contrast to, e.g., "2-4" potential. Fortunately in our case the eigenfunctions of SE are known (\ref{eq16}) that provides an alternative direct way for obtaining  $K$. From \cite{Col85} we have the expressions
\begin{equation}
\label{eq31}  <\bar d \mid \exp \left(-HT/\hbar \right) \mid -\bar d> =\sqrt{ \frac{\omega}{\pi \hbar}}\ e^{-\omega T/2}
 sh\left[KT\exp \left(-S_0/\hbar \right)\right]
\end{equation}
\begin{equation}
\label{eq32}  <-\bar d \mid \exp \left(-HT/\hbar \right) \mid -\bar d> =\sqrt{ \frac{\omega}{\pi \hbar}}\ e^{-\omega T/2}
 ch\left[KT\exp \left(-S_0/\hbar \right)\right]
\end{equation}
where $H$ is the hamiltonian of our SE (\ref{eq1}).
From here we express $K$ as
\begin{equation}
\label{eq33} K=\frac{1}{T}\exp \left(S_0/\hbar \right)\ arcth\left(\frac{<\bar d \mid \exp \left(-HT/\hbar \right) \mid -\bar d>}{ <-\bar d \mid \exp \left(-HT/\hbar \right) \mid -\bar d>}\right)
\end{equation}
Substitution of (\ref{eq33}) into (\ref{eq27}) yields
\begin{equation}
\label{eq34}  E_1-E_0=\frac{2\hbar}{T}\left(\frac{S_0}{2\pi\hbar}\right)^{1/2}\ arcth\left(\frac{<\bar d \mid \exp \left(-HT/\hbar \right) \mid -\bar d>}{ <-\bar d \mid \exp \left(-HT/\hbar \right) \mid -\bar d>}\right)
\end{equation}
Taking into account our eigenfunctions $\mid q>\equiv\psi_q (x)$ given by (\ref{eq16}) with dimensional coordinate $x$ related to the dimensionless $y$ by (\ref{eq2}) we have
\begin{equation}
\label{eq35}  \Biggl <\bar d \left \vert \exp \left(-\frac{HT}{\hbar} \right) \right \vert -\bar d \Biggr> =\sum_{n=0}^{\infty}\ e^{-\frac{\left(E_n-V\left(\bar d\right)\right) T}{\hbar} } <\bar d \mid n><n \mid -\bar d>
\end{equation}
\begin{equation}
\label{eq36} \Biggl <-\bar d \left \vert \exp \left(-\frac{HT}{\hbar} \right) \right \vert -\bar d \Biggr> =\sum_{n=0}^{\infty}\ e^{-\frac{\left(E_n-V\left(\bar d\right)\right) T}{\hbar} }  <-\bar d \mid n><n \mid -\bar d>
\end{equation}
The wave functions in the positions of the potential minima are $\mid \bar d>=\delta \left(x-\bar d \right)$ and $\mid -\bar d>=\delta \left(x+\bar d \right)$ respectively.

To proceed further we need to know $T$. From now on we will use dimensionless values. Introducing the dimensionless time as
\begin{equation}
\label{eq37}  s=\tau\frac{\pi^2\hbar}{4ML^2}
\end{equation}
we rewrite (\ref{eq30}) in dimensionless units
\begin{equation}
\label{eq38} s=\int_{0}^{y_{cl}}\frac{dy}{\sqrt{U(y)-U\left( d\right)}}
\end{equation}
Introducing the dimensionless analog of $T$ by the requirement $y_{cl}(S/2)=d$ we obtain for $S$ the following relationship
\begin{equation}
\label{eq39} S=2\int_{0}^{d}\frac{dy}{\sqrt{U(y)-U\left( d\right)}}
\end{equation}
Its solution for malonaldehyde is $S=0.5366$. As a result we have
\begin{equation}
\label{eq40}  \Biggl <\bar d \left \vert \exp \left(-\frac{HT}{\hbar} \right) \right \vert -\bar d \Biggr> =\sum_{n=0}^{\infty}\ e^{-\left(\epsilon_n-U\left( d\right)\right) S/2 } \psi_n (d)\psi_n (-d)
\end{equation}
\begin{equation}
\label{eq41}  \Biggl <-\bar d \left \vert \exp \left(-\frac{HT}{\hbar} \right) \right \vert -\bar d \Biggr> =\sum_{n=0}^{\infty}\ e^{-\left(\epsilon_n-U\left( d\right)\right) S/2 } \left[\psi_n (-d)\right]^2
\end{equation}
Finally we obtain from (\ref{eq34})
\[
\epsilon_1-\epsilon_0=\frac{4}{S}\left(\frac{1}{2\pi}\int_{-d}^{d}dy\ \sqrt{U(y)-U(d)}\right)^{1/2}\times
\]
\begin{equation}
\label{eq42} arcth \left\{\frac{\sum_{n=0}^{\infty}\ e^{-\left(\epsilon_n-U\left( d\right)\right) S/2 } \psi_n (d)\psi_n (-d)}{\sum_{n=0}^{\infty}\ e^{-\left(\epsilon_n-U\left( d\right)\right) S/2 } \left[\psi_n (-d)\right]^2}\right\}
\end{equation}
Taking into account only four lowest energy levels (i.e., approximating $\infty$ by $3$ in the sums) we obtain for malonaldehyde $\epsilon_1-\epsilon_0=0.839335$.

\section{Results and discussion}
Fig.1 shows that the parameters of the potential (\ref{eq3}) can be chosen to provide good description of the energy levels structure for a set of specific experimental data. In Fig.1 the energy levels for the hydrogen bond in malonaldehyde (experimental data determined with help of the IR spectroscopy are taken from \cite{Fil05}) are presented. The energy levels for the proton in the hydrogen bond form a doublet within the wells and the second one around the barrier top. The ground-state splitting is $E_1-E_0=21.583\ {\rm cm^{-1}}$. The transition frequencies for the upper states are $E_2-E_0=2450\ {\rm cm^{-1}}$ and $E_3-E_0=2960\ {\rm cm^{-1}}$. These experimental values are obtained from our dimensionless ones if we take $m=84$ ($h\approx 7055.75$), $b=9514.78$ ($p\approx 97.544$), $a=0$. For malonaldehyde the distance between oxygen atoms is $2.58\ \AA$ (obtained from {\it ab initio} quantum chemical calculations \cite{Yan10}) so that $L=1.29\ \AA$. We also take into account that for hydrogen $m_H=1\ {\rm amu}$ and for oxygen $m_O=16\ {\rm amu}$ so that a fictitious quantum particle with the reduced mass (the associated reduced mass of the system along the coordinate $x$) is
\begin{equation}
\label{eq43} M=\frac{2m_Hm_O}{m_H+2m_O}\approx 0.97
\end{equation}
It practically coincides with the proton mass that validates our model with fixed positions of oxygen atoms (constant $\pm L$ values).

The exact result for the ground state splitting in dimensionless units is $\epsilon_1-\epsilon_0=0.909$. The results for different variants of WKB approach are compared with this referee point as follows: 1. Landau and Lifshitz textbook formula $\epsilon_1-\epsilon_0=0.922466$; 2. Garg's formula $\epsilon_1-\epsilon_0=0.973715$; 3. instanton approach $\epsilon_1-\epsilon_0=0.839335$.

One can see that the formula from the Landau and Lifshitz textbook (ordinary WKB method) provides the best accuracy of all three investigated WKB variants. However it should be stressed that this formula works well only if the necessary input information is available (the turning points corresponding to $E_m=\left(E_1-E_0\right)/2$). These turning points a priory are unknown and its not clear how one can obtain them for an arbitrary DWP of interest (other than the considered trigonometric one) except by numerical solution of the corresponding SE. In our case of trigonometric DWP this information is known as a result of the exact solution of SE. In the general case of an arbitrary DWP it can be obtained only as a result of tedious numerical calculations. But numerical solution of SE can give the ground state splitting directly and makes WKB estimate to be redundant. Thus the utility of the formula from the Landau and Lifshitz textbook is extremely limited.

Garg's formula \cite{Gar00} for a symmetric potential (\ref{eq22}) yields more crude estimate of the ground state splitting than that (\ref{eq17}) from the Landau and Lifshitz textbook  \cite{Lan74}. However the (\ref{eq22}) indeed have a considerable advantage (noted in \cite{Gar00}) compared with (\ref{eq17}). In the latter the integration is carried out between the turning points corresponding to $E_m$. In contrast in (\ref{eq22}) the integration is carried out between the minima the potential that are known from the shape of DWP. We conclude that Garg's formula \cite{Gar00} both provides satisfactory accuracy and is very convenient for usage.

Instanton approach was initially conceived for the usage in the quantum field theory \cite{Col85}, \cite{Kle95}. Its application to the problem of tunneling is always exposed as a somewhat toy exercise. However even in the most elaborate cases ("$2-4$" potential \cite{Col85}, \cite{Kle95}, \cite{Gil77} or the "pendulum $1-cos$" one \cite{Neu78}) it produces very cumbersome formulas that are difficult for application. For our trigonometric DWP such calculations are impossible but the knowledge of the exact solution of SE enables us to circumvent the difficulties in an alternative way. The result is also less accurate than that of the formula from the Landau and Lifshitz textbook and the calculations are rather tedious. Taking into account higher levels does not improve the accuracy of the instanton approach. For instance taking six lowest energy levels (i.e., approximating $\infty$ by $5$ in the sums) does not change the value $\epsilon_1-\epsilon_0=0.839335$ at all. Garg showed that (\ref{eq22}) is equivalent to the formula given by the instanton approach \cite{Col85}, \cite{Kle95}. However it was shown within the framework of Coleman's approximation
\begin{equation}
\label{eq44} K\approx \sqrt{2\omega}\beta
\end{equation}
where $\beta$ is given by the asymptotic behavior $\tau\rightarrow \infty$ of the instanton velocity 
\begin{equation}
\label{eq45} \left(\frac{M}{S_0}\right)^{1/2}\frac{dx_{cl}}{d\tau}\approx \beta \exp \left(\omega \tau \right)
\end{equation}
On our approach we do not use this approximation. As a result we obtain that the estimate based on the Garg's formula differs from that of instanton approach. Both results are roughly similarly inaccurate compared with that of the ordinary WKB. We conclude that Garg's formula provides similar accuracy as the instanton approach but is much more convenient for applications.

There is a well known difference between the ordinary WKB and instanton calculations for the ground state splitting \cite{Gil77}, \cite{Neu78}
\begin{equation}
\label{eq46} \frac{\Delta E_{WKB}}{\Delta E_{instanton}}=\left(\frac{e}{\pi}\right)^{1/2}\approx 0.93
\end{equation}
This difference is obtained for the "$2-4$" potential \cite{Gil77} and for the "pendulum $1-cos$" one \cite{Neu78}. However in the general case this difference is essentially potential dependent. For instance in our case of the trigonometric DWP the instanton result is $0.839335$ and we have $0.839335*0.93=0.780582$ that is not at all $0.922466$ given by ordinary WKB method.

One can conclude that the Schr\"odinger equation with symmetric trigonometric double-well potential can be exactly solved via angular oblate spheroidal function. Our stringent analytic description of the ground state splitting can well be a referee point for comparison of the accuracy of numerous WKB formulas suggested in the literature. We conclude that if the necessary input information for the formula from the Landau and Lifshitz textbook is available (as in the case of our trigonometric DWP) then the latter provides more accurate result that both the Garg's formula or that from instanton approach. The exact solution well suits for the hydrogen bond in malonaldehyde. Thus it yields a new theoretical tool for the description of this important molecule. The results obtained provide good quantitative description of relevant experimental data on IR spectroscopy of malonaldehyde.

Acknowledgements. The author is grateful to Dr. Yu.F. Zuev
for helpful discussions. The work was supported by the grant from RFBR N 15-29-01239.

\end{document}